\documentclass[
  reprint,
  aps,
  pre,
  superscriptaddress
]{revtex4-2}

\usepackage{tabularx}
\usepackage{color}
\usepackage{mathtools}
\usepackage{hyperref}
\usepackage{amsmath,amssymb}
\usepackage{graphicx}
\usepackage{dcolumn}
\usepackage{bm}
\usepackage{float}
\usepackage{epstopdf}
\usepackage{textcomp}
\usepackage[shortlabels]{enumitem}
\usepackage{verbatim}

\def\be{\begin{equation}}
	\def\ee{\end{equation}}
\def\bea{\begin{eqnarray}}
	\def\eea{\end{eqnarray}}

\newcommand{\rev}[1]{#1}

\newcommand{\cmcb}{\affiliation{Centre for Mechanochemical Cell Biology and Division of Biomedical Sciences, Warwick Medical School, University of Warwick, Coventry CV4 7AL, United Kingdom.}}

\begin{document}
    \title{Confinement Limits Cell Shape and Mitotic Pressure During Division}
	\author{Amit Singh Vishen}
    \email{amit.singh-vishen@warwick.ac.uk}
	\cmcb
	\date{November 10, 2026}
	
\begin{abstract}
Cells often divide under mechanical confinement, where surrounding structures restrict shape changes during cytokinesis. We analyze a minimal model of a dividing cell as an incompressible volume bounded by an interface with effective isotropic tension, and show that confinement imposes a minimum attainable furrow radius. Beyond this limit, the cytokinetic ring can continue to contract within a septum without further furrow ingression, leaving the cell shape, mitotic pressure, and transmitted force unchanged. We determine this confinement-limited state for rigid parallel walls, soft axial confinement, and three-dimensional enclosures. Between parallel walls, the limiting force and minimum furrow radius collapse onto universal curves when scaled by surface tension and cell size. In soft confinement, environmental stiffness and spindle forces select the operating point. In three-dimensional confinement, the free volume between the cell and the confinement controls force transmission and cell shape. This yields a method for inferring the mechanical state of a cell dividing in deformable three-dimensional confinement from its shape. The radius and tangent angle of the cell surface at the stalled furrow and at the point where the cell detaches from its surroundings determine the mitotic pressure and the axial force carried by the cell surface, relative to the cortical tension.
\end{abstract}

\maketitle

\section{Introduction}

Cell division is an intrinsically mechanical process in which internally generated forces reorganize cell shape and partition volume into two daughter compartments \cite{Guillot2013,Scholey2003,Pollard2019}. In physiological settings, division occurs in mechanical contact with neighboring cells, tissues, or enclosing structures. The forces generated during division drive tissue morphogenesis and determine whether cells can successfully complete cytokinesis in mechanically restrictive environments such as dense tissues or narrow tissue spaces \cite{Cattin2015, Campinho2013, Wang2017, Tang2018, Tarannum2022}. 
Understanding how confinement restricts cell shape during division, and how these  constraints influence mitotic pressure and force transmission, is therefore important for understanding tissue organization, developmental mechanics, and division failure in confined environments.

The mechanical aspects of mitosis have been studied extensively, particularly during mitotic rounding, when cells increase surface tension to regain a nearly spherical shape. Experimental and theoretical work has quantified the pressures associated with rounding and related them to cortical tension and hydrostatic pressure \cite{Stewart2011, Ramanathan2015, Fischer2014, Cattin2015, Sorce2015, Lancaster2013, Taubenberger2020, Yang2021}. 
Recent experiments show that dividing cells also exert substantial forces during anaphase and cytokinesis, often exceeding those generated during rounding \cite{Nam2018, Gupta2021, Nam2021}. For single cells in three-dimensional alginate gels, protrusive stresses during anaphase~B and cytokinesis were estimated at $1$--$2\,{\rm kPa}$, compared with at most $\sim 0.8\,{\rm kPa}$ during mitotic swelling \cite{Nam2018}. In collagen gels, dividing cells displace the matrix by $\sim$1\,$\mu$m along the division axis, and in epithelial monolayers on soft substrates, division-elongation forces were estimated at $300$--$500\,{\rm nN}$ \cite{Gupta2021, Nam2021}, compared with $50$--$100\,{\rm nN}$ for mitotic rounding \cite{Cattin2015, Stewart2011}.
Perturbation experiments identify contraction of the cytokinetic ring as a main source of these forces. Because cell volume is nearly conserved through division, lateral contraction at the equator must be accommodated by expansion along the division axis, an argument supported by simulations of membrane-bounded cells \cite{Nam2018, Gupta2021, Nam2021}. How this conversion translates into mitotic pressure and transmitted force, and how the confinement geometry bounds it, has not been quantified.

The difficulty is geometric: the contractile ring generates circumferential tension and contracts radially at the division plane, whereas the forces transmitted to the confining boundary act along the division axis.
A dividing cell can accommodate radial contraction through shape changes, curvature redistribution, and axial extension. Previous theoretical approaches have modeled cytokinesis using active gel descriptions of the cortex and contractile ring dynamics \cite{Turlier2014,Salbreux2017,Dorn2016,Chatterjee2022}, but without confinement. Simulations accompanying recent force measurements \cite{Nam2018,Gupta2021,Nam2021,Rabbi2026} include the environment, but represent the cell surface as an elastic membrane and impose ring constriction as displacements of equatorial membrane nodes. In these simulations the transmitted force depends on membrane elastic parameters, and the ring and the furrow cannot separate. What is lacking is a framework that relates the confinement geometry to the cell shapes attainable during division, and hence to the transmitted force and mitotic pressure.

Here we ask, how does confinement geometry restrict the cell shapes that can be attained during division, and what consequences does this have for mitotic pressure and force transmission? To address these questions, we model the dividing cell as an incompressible fluid enclosed by a membrane–cortex interface with constant effective surface tension and analyze force balance under geometric constraints. This description is closely analogous to classical capillary-bridge problems, in which  shapes are determined by geometry, incompressibility, and surface tension \cite{Yoneda1972, Brochard2003, Fischer2014, Yang2017, Yang2021, Vishen2024}.

We analyze three forms of confinement. First, rigid parallel-wall confinement isolates the geometric constraint imposed on cell shape and establishes the force--confinement relation. Second, soft confinement introduces environmental compliance and describes how tissue stiffness determines the force and pressure exerted by the dividing cell; forces generated by the mitotic spindle are included as additional axial driving terms in the force balance. Third, cylindrical confinement considers a three-dimensional enclosure, where the available free volume between the cell and the confining boundary determines how much deformation can be accommodated before force is transmitted to the boundary. Finally, we show that the mechanical state of a dividing cell can be inferred from its shape alone, discuss the effect of adhesion to the environment, and compare the resulting scales with experiments.

\begin{figure}
    \centering
  \includegraphics[width=0.9\linewidth]{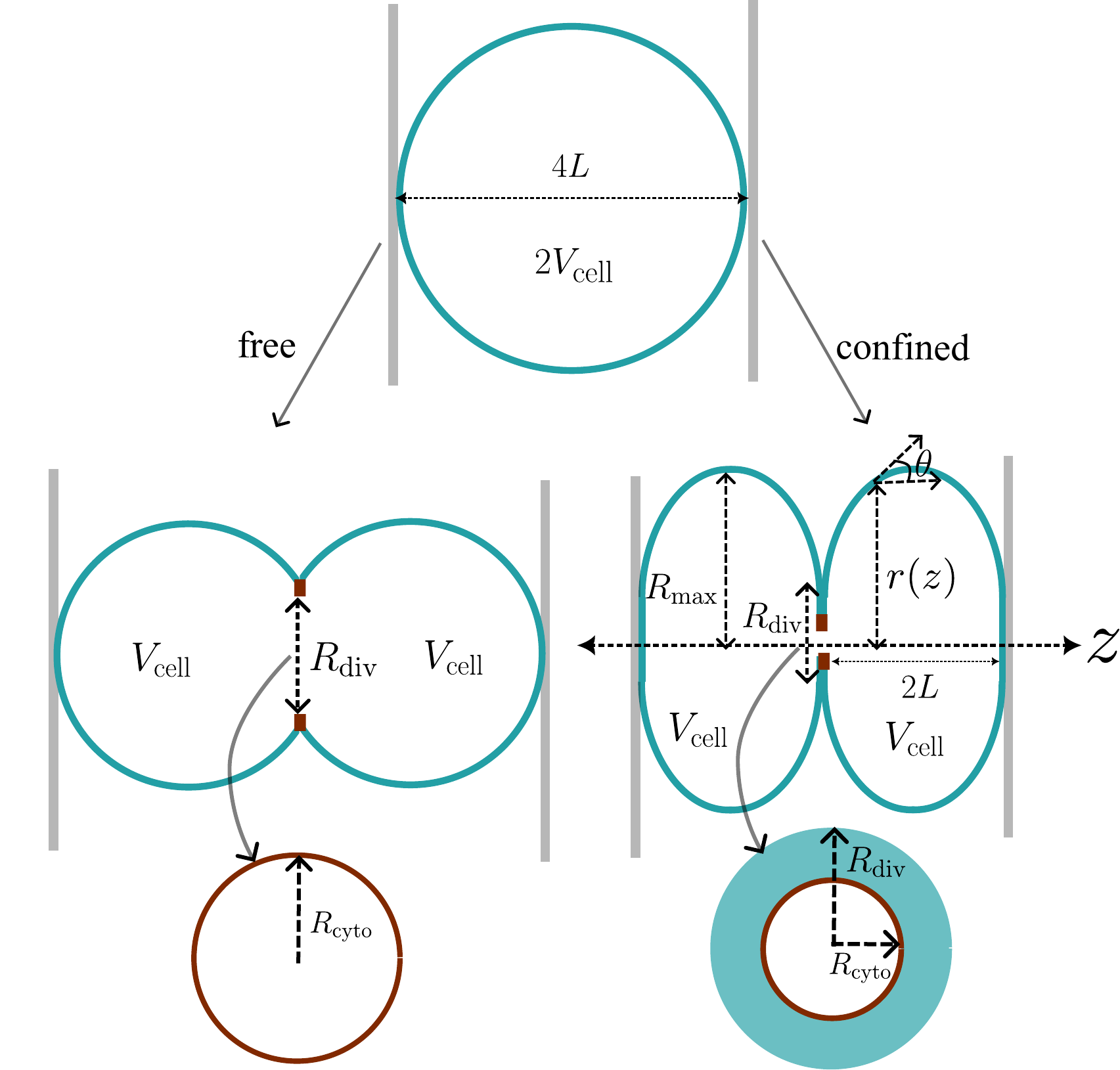}
\caption{Schematic of cell division without confinement (free) and under rigid parallel-wall confinement (confined). Top: the undivided cell of fixed volume $2V_\mathrm{cell}$, shown just touching walls separated by $4L$. The cell is characterized by the geometric furrow radius $R_\mathrm{div}$ at the division plane and by its local radius $r(z)$, where $\theta$ is the angle between the surface tangent and the division axis $z$, and $R_\mathrm{max}$ is the maximum radius. Without confinement, the furrow follows the cytokinetic ring ($R_\mathrm{div}=R_\mathrm{cyto}$). Under confinement, axial elongation is restricted to $2L$ per daughter of volume $V_\mathrm{cell}$. Once the furrow reaches its minimum radius $R_\mathrm{div}^\mathrm{min}$, the ring (red) continues to constrict within a septum (shaded) of outer radius $R_\mathrm{div}^\mathrm{min}$. Bottom: the division plane, showing the ring radius $R_\mathrm{cyto}$ and the furrow radius $R_\mathrm{div}$.}
    \label{fig1}
\end{figure}

\begin{table}[t]
\centering
\caption{Dimensionless variables used for axial and three-dimensional confinement.}
\label{tab:notation}
\begin{tabular}{lcc}
\hline
\noalign{\vskip 4pt}
\textbf{Quantity} & \textbf{Axial} & \textbf{3D} \\
\noalign{\vskip 4pt}
\hline
\noalign{\vskip 4pt}

Length scale
& $R_{\mathrm{cell}}$
& $R_{\mathrm{cyl}}$ \\[5pt]

Force
& $\displaystyle f=\frac{F}{2\pi\gamma R_{\mathrm{cell}}}$
& $\displaystyle \tilde f=\frac{F}{2\pi\gamma R_{\mathrm{cyl}}}$ \\[7pt]

Pressure
& $\displaystyle p=\frac{\Delta P}{2\gamma/R_{\mathrm{cell}}}$
& $\displaystyle \tilde p=\frac{\Delta P}{2\gamma/R_{\mathrm{cyl}}}$ \\[7pt]

Radius
& $\displaystyle \rho=\frac{r}{R_{\mathrm{cell}}}$
& $\displaystyle \tilde\rho=\frac{r}{R_{\mathrm{cyl}}}$ \\[7pt]

Confinement parameter
& $\displaystyle l=\frac{L}{R_{\mathrm{cell}}}$
& $\displaystyle v=\frac{\Delta V}{\Delta V_{\mathrm{th}}}$ \\

\noalign{\vskip 4pt}
\hline
\end{tabular}
\end{table}

\section{Rigid axial confinement}
\label{sec:rigid}

We begin by analyzing cell division under rigid confinement in order to establish the mechanical and geometric framework used throughout this work. We consider a dividing animal cell of fixed total volume \(2V_{\rm cell}=8\pi R_{\rm cell}^3/3\), where \(R_{\rm cell}\) is the radius of a sphere having the volume of one daughter cell. The cell is confined between two parallel, rigid walls separated by a fixed distance $4L$ (Fig.\,\ref{fig1}). The cell is modeled as an incompressible fluid enclosed by a membrane-cortex interface with constant effective surface tension $\gamma$. 

A key distinction maintained throughout this work is between the cytokinetic ring radius $R_\mathrm{cyto}$ and the geometric furrow radius $R_\mathrm{div}$, defined as the radius of the membrane-cortex interface at the division plane. As we show below, while $R_\mathrm{cyto}$ may decrease continuously during cytokinesis and can in principle approach zero, 
the furrow radius $R_\mathrm{div}$ is constrained by confinement. It need not follow $R_\mathrm{cyto}$ once the confinement-induced minimum furrow radius is reached (Fig.~1).

The cell shape is assumed to be axisymmetric about the division axis $z$ and is described by a local radius $r(z)$. Mechanical equilibrium requires force balance across any cross-section perpendicular to the division axis. In the presence of a compressive force $F$ exerted by the confining walls, this balance reads \cite{Brochard2002,Brochard2003,Vishen2024}
\begin{equation}
\label{eq:forcebalance1}
2\pi r \gamma \cos\theta - \pi r^2 \Delta P + F = 0,
\end{equation}
where $\theta$ is the angle between the membrane tangent and the division axis. Positive $F$ corresponds to compression by the walls. 
The pressure difference is related to the curvature through the Young-Laplace law: $\Delta P = 2\gamma/R_\mathrm{H}$, where $R_\mathrm{H}$ is the mean radius of curvature. 

We nondimensionalize lengths by $R_{\rm cell}$, forces by
$F_0=2\pi\gamma R_{\rm cell}$, and pressure by
$\Delta P_0=2\gamma/R_{\rm cell}$. We denote the dimensionless local radius, force, and pressure by
$\rho=r/R_{\rm cell}$, $f=F/F_0$, and $p=\Delta P/\Delta P_0$, respectively (Table \ref{tab:notation}). In terms of these variables, Eq.~\eqref{eq:forcebalance1} becomes
\begin{equation}
\label{eq:forcebalance2}
\rho^2-\frac{\cos\theta}{p}\rho-\frac{f}{p}=0.
\end{equation}
The two solutions of $\rho$ are
\begin{equation}
\label{eq:radius2}\rho_{\pm}
=
\frac{\cos\theta\pm
\sqrt{\cos^2\theta+4pf}}{2p}.
\end{equation}
For compressive confinement ($f>0$), only the positive branch corresponds to a physical solution. The cell shape can then be obtained by parametrizing the interface by arc length $s$ and imposing the appropriate boundary conditions (see Appendix~\ref{appendix_shape}).

\begin{figure}
    \centering
  \includegraphics[width=\linewidth]{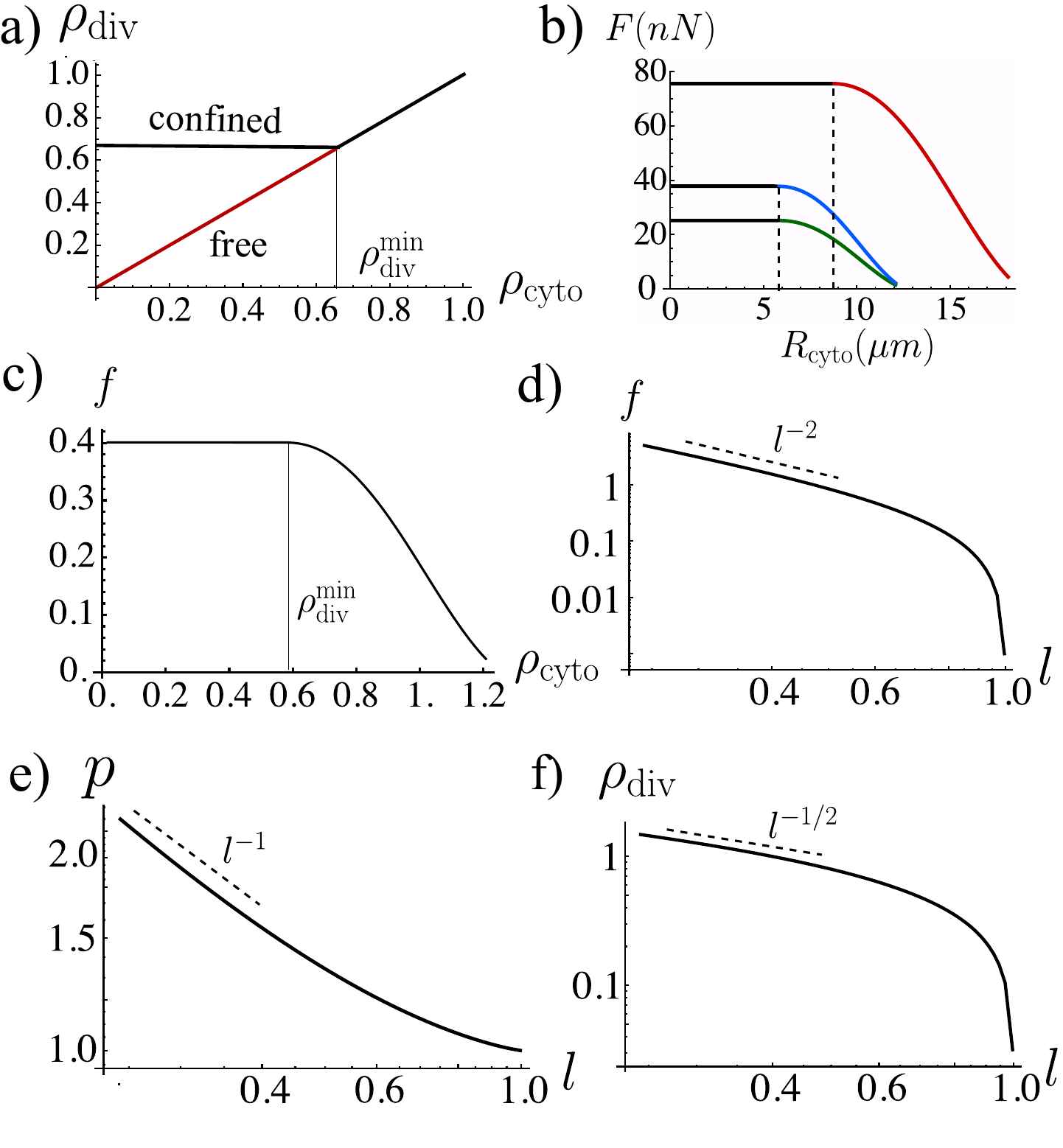}
\caption{\textbf{Cell division under rigid axial confinement.}
(a) Scaled geometric furrow radius $\rho_\mathrm{div}$ as a function of the cytokinetic-ring radius $\rho_\mathrm{cyto}$ at $l=2^{-2/3}$. Without confinement (red), $\rho_\mathrm{div}=\rho_\mathrm{cyto}$. Under confinement (black), $\rho_\mathrm{div}$ saturates at the confinement-induced minimum $\rho_\mathrm{div}^\mathrm{min}$. 
(b) Axial force in physical units as a function of $R_\mathrm{cyto}$ at fixed confinement $l=2^{-2/3}$, for which the undivided cell just touches the walls ($L=2^{-2/3}R_\mathrm{cell}$). Curves are shown for $(\gamma,R_b) =  (1\,\mathrm{mN/m},12\,\mu\mathrm{m})$ (green), $(1.5\,\mathrm{mN/m},12\,\mu\mathrm{m})$ (blue) and
$(1.5\,\mathrm{mN/m},18\,\mu\mathrm{m})$ (red), where $R_b=2^{1/3}R_\mathrm{cell}$ is the radius of the undivided cell. These correspond to $R_\mathrm{cell}\approx 9.5\,\mu\mathrm{m}$, $L=6\,\mu\mathrm{m}$ (green, blue) and $R_\mathrm{cell}\approx 14.3\,\mu\mathrm{m}$, $L=9\,\mu\mathrm{m}$ (red). Dashed vertical lines mark $R_\mathrm{div}^\mathrm{min}$. Increasing $\gamma$ at fixed $R_\mathrm{cell}$ rescales the force without changing $R_\mathrm{div}^\mathrm{min}$ (green and blue). Increasing $R_\mathrm{cell}$ at fixed $l$ increases both $R_\mathrm{div}^\mathrm{min}\propto R_\mathrm{cell}$ and the saturating force $\propto\gamma R_\mathrm{cell}$ (blue and red).
(c) Dimensionless axial force $f$ as a function of $\rho_\mathrm{cyto}$ at $l=2^{-2/3}$. The force increases during furrow ingression and saturates once $\rho_\mathrm{div}^\mathrm{min}$ is reached. Curves for different $\gamma$ and $R_\mathrm{cell}$ at the same $l$ collapse onto this curve.
(d) Saturation force $f$, (e) saturation pressure difference $p$, and
(f) minimum furrow radius $\rho_\mathrm{div}^\mathrm{min}$ as functions of the dimensionless confinement length $l=L/R_\mathrm{cell}$ (wall separation $4L$) at fixed cell volume, on logarithmic axes. Dashed lines show the strong-confinement scalings $f\sim l^{-2}$, $p\sim l^{-1}$ and $\rho_\mathrm{div}^\mathrm{min}\sim l^{-1/2}$. Except in (b), lengths are scaled by $R_\mathrm{cell}$, forces by $F_0=2\pi\gamma R_\mathrm{cell}$, and pressure differences by $\Delta P_0=2\gamma/R_\mathrm{cell}$.}
    \label{fig2}
\end{figure}

The maximum cell radius occurs where the surface tangent is parallel to the division axis ($\theta = 0$). At the division plane, the tangent angle is controlled by the cytokinetic ring. Since $\rho_+$ increases monotonically with $\cos\theta$, the furrow radius decreases as $\theta$ increases at the division plane. Radii smaller than the value at $\theta = \pi/2$ would require $\theta > \pi/2$, for which the interface approaching the division plane would first cross it. The minimum furrow radius is therefore attained when the surface tangent is perpendicular to the division axis ($\theta = \pi/2$). From Eq.~(\ref{eq:radius2}), the confinement-imposed minimum furrow radius is then
\begin{equation}
  \rho_\mathrm{div}^\mathrm{min} = \sqrt{f/p}.
  \label{eq:min-rho}
\end{equation}
This is the first main result: the radius at the division plane has a minimum value.  In dimensionless form, the furrow radius therefore satisfies
\begin{equation}
\label{eq:radius-div}
\rho_{\rm div} =
\max\!\left(\rho_{\rm cyto},\; \rho_{\rm div}^{\min}\right),
\end{equation}
where $\rho_{\rm div}=R_{\rm div}/R_{\rm cell}$ and $\rho_{\rm cyto}=R_{\rm cyto}/R_{\rm cell}$. For $\rho_{\rm cyto}>\rho_{\rm div}^{\min}$, the interface follows ring constriction and $\rho_{\rm div}\simeq\rho_{\rm cyto}$. Once $\rho_{\rm cyto}$ reaches $\rho_{\rm div}^{\min}$, further ring contraction does not lead to additional furrow ingression (Fig.~\ref{fig2}(a)). 
The corresponding axial force balance can be generalized to an axisymmetric interface with spatially varying surface stresses. This generalization, and its implication for the minimum furrow radius, are discussed in Appendix~\ref{appendix_general}.
We have assumed that each membrane of the septum has the same tension $\gamma$ as the outer surface. A lower tension $\gamma_s<\gamma$, for example from adhesion with energy $w$ per unit area ($\gamma_s=\gamma-w/2$), sets the contact angle through $\sin\theta_c=\gamma_s/\gamma$, and Eq.~(3) with $\theta=\theta_c$ then gives a larger minimum furrow radius.

\rev{
To construct the quasistatic sequence of cell shapes during cytokinesis, we fix the cell volume at $V_{\rm cell}/R_{\rm cell}^3=4\pi/3$ and specify the confinement through the dimensionless confinement length $l=L/R_{\rm cell}$. For each value of the prescribed cytokinetic-ring radius $\rho_{\rm cyto}$, the interface shape is determined from Eq.~\eqref{eq:radius2} with the appropriate boundary conditions, while the force $f$ and pressure $p$ are determined by the confinement and volume constraints (Appendix~\ref{appendix_shape}). 
Varying $\rho_{\rm cyto}$ at fixed $l$ therefore gives the quasistatic force--ring-radius and furrow-radius relations shown in Fig.~\ref{fig2}(a,b). Once $\rho_{\rm cyto}$ reaches $\rho_{\rm div}^{\min}$, further ring contraction does not change the interface geometry, and the force and pressure remain at their confinement-limited values, shown in Fig.~\ref{fig2}(b).
}

\rev{Fig.~2(b) illustrates the distinct roles of cell size and surface tension. Varying the cell size shifts both the minimum attainable furrow radius and the saturating force, whereas varying the surface tension rescales the saturating force while leaving the minimum furrow radius unchanged. As seen from Fig.~2(c), after scaling the force by $F_0$ and the radius by $R_\mathrm{cell}$, the curves collapse.}

\rev{
We note that this construction separates the mechanical response from the dynamics of ring contraction. A dynamical model could specify an evolution law $\rho_{\rm cyto}(t)$, in which case the quasistatic solution would give $f(t)=f[\rho_{\rm cyto}(t)]$ and $p(t)=p[\rho_{\rm cyto}(t)]$. 
A minimal dynamical description of this decoupling between ring contraction and furrow ingression is presented in Appendix~\ref{appendix_ring}.}

\rev{We next characterize the confinement-limited state by setting $\rho_{\rm cyto}=\rho_{\rm div}^{\min}$ and varying the dimensionless confinement $l$. In this limit, the dimensionless cell shape, force, and pressure are determined by the single geometric parameter $l$, since the dimensionless  volume is fixed at $4\pi/3$. }
\rev{The scaled saturation force and pressure are shown in Figs.~\ref{fig2}(d) and \ref{fig2}(e), respectively, and the corresponding scaled minimum furrow radius is shown in Fig.~\ref{fig2}(f). 
In the strong-confinement regime, the confinement length sets the characteristic meridional curvature, giving $R_{\rm H}/R_{\rm cell}\sim l$ and hence $ p \sim l^{-1}$. Fixed cell volume then requires the transverse dimension, and therefore the minimum furrow radius, to scale as $\rho_{\rm div} \sim l^{-1/2}$. At the limiting furrow, Eq.~(\ref{eq:min-rho}) consequently gives $f \sim l^{-2}$. Figs.~\ref{fig2}(d-f) show the scaling with $l$ overlaid on the full curve.  }

\section{Soft Axial Confinement}

\begin{figure}
    \centering
  \includegraphics[width=\linewidth]{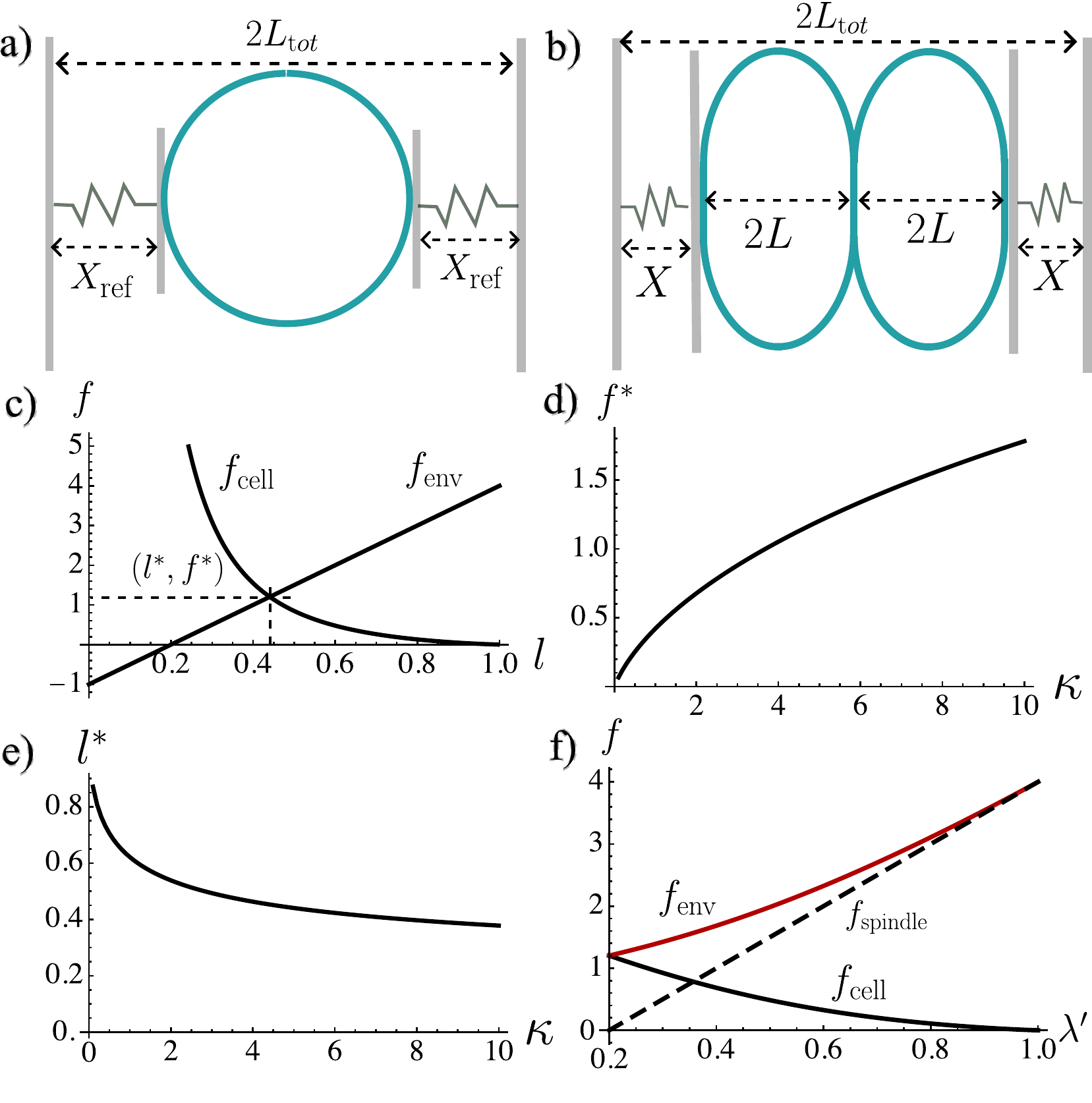}
\caption{\textbf{Cell division under soft confinement.}
(a--b) Schematic of soft axial confinement modeled as an effective linear elastic element of stiffness $k$ and reference length $X_\mathrm{ref}$ on each side of the cell. The outer boundaries are separated by $2L_\mathrm{tot}$. During cytokinesis, furrow ingression elongates the cell along the division axis (daughter length $2L$) and compresses the surrounding medium to length $X=L_\mathrm{tot}-2L$.
(c) Graphical construction of mechanical equilibrium for $\kappa=5$ and $\lambda=0.2$. The axial force generated by the cell under rigid confinement, $f_\mathrm{cell}$, is shown together with the elastic force--length relation of the environment, $f_\mathrm{env}$. The operating point $(l^*,f^*)$, where $f_\mathrm{cell}=f_\mathrm{env}$, is given by the intersection of the two curves. Increasing stiffness $\kappa$ steepens the environmental response and shifts the equilibrium toward smaller separations and larger forces. In the limit $\kappa\to\infty$, the solution approaches the rigid-wall limit with $l=\lambda$.
(d) Equilibrium force $f^*$ and (e) equilibrium daughter half-length $l^*$ as functions of the dimensionless stiffness $\kappa$, for $\lambda=0.2$.
(f) Effect of a constant spindle force at fixed $\kappa=5$ and $\lambda=0.2$, plotted against $\lambda'\equiv\lambda+f_\mathrm{spindle}/\kappa$. The curves show the cell-surface force $f_\mathrm{cell}=f^*$ (black), the spindle force $f_\mathrm{spindle}=\kappa(\lambda'-\lambda)$ (dashed), and the total force on the environment $f_\mathrm{env}=f_\mathrm{cell}+f_\mathrm{spindle}$ (red). Increasing the spindle force lowers $f^*$ while raising $f_\mathrm{env}$. Forces are scaled by $F_0=2\pi\gamma R_\mathrm{cell}$ and lengths by $R_\mathrm{cell}$.}
 
    \label{fig3}
\end{figure}

So far we have considered confinement by rigid walls, for which the cell half-length $L$ is prescribed. However, in many biological contexts, such as a cell dividing in a tissue, the surrounding medium is deformable. We model the soft confining environment as a linear elastic element  with effective stiffness $k$ and reference length $X_{\rm ref}$ (Fig.~\ref{fig3}(a)).
\rev{We define the total half-separation between the two outer boundaries as $L_{\rm tot}$. If the cells after division have half-length $L$, the remaining length available to the confining environment is $X=L_{\rm tot}-2L$ (Fig.~\ref{fig3}(b)). The force on the cell from the environment is $F_{\rm env} = -k (X - X_{\rm ref})$, which in dimensionless form reads:
\begin{equation}
f_{\rm env}(l)
=
\kappa\left(l-\lambda\right),
\label{eq:f-env}
\end{equation}
where we have defined $f_{\rm env} = F_{\rm env}/2\pi \gamma R_{\rm cell}$, $\kappa=k/\pi\gamma, \lambda= (L_{\rm tot}-X_{\rm ref})/2R_{\rm cell}$. $\kappa$ measures the stiffness of the environment relative to the cellular surface-tension and $\lambda$ is the dimensionless zero-force confinement length.\\
The rigid-confinement calculation in Sec.~II provides a constitutive relation $f_{\rm cell}(l)$, giving the force exerted by the cell surface for each imposed confinement length (Fig.~2(d)). Mechanical equilibrium under soft confinement is therefore obtained by intersecting this relation with the force--length relation of the environment as given by Eq.~(\ref{eq:f-env}),
\begin{equation}
f_{\rm cell}(l^\ast)=f_{\rm env}(l^\ast),
\end{equation}
which determines the equilibrium cell length $l^\ast$ and force $f^\ast$.}

\rev{As shown in Fig.~\ref{fig3}(c), the soft-confinement solution is given by the intersection of the rigid-cell force--confinement relation $f_{\rm cell}(l)$ with a straight line of slope $\kappa$ and zero-force intercept $\lambda$.}
Increasing $\kappa$ increases the slope of the environmental force-length relation, shifting the operating point toward the rigid-confinement limit. In the limit $\kappa\rightarrow\infty$, the equilibrium approaches the rigid-wall solution with confinement $l=\lambda$. 
In contrast, changing $\lambda$ translates the environmental force-length relation and therefore changes the degree of confinement at fixed environmental stiffness (Fig.~3(c)). The corresponding equilibrium force and daughter cell half-length are shown as functions of $\kappa$ in Figs.~3(d) and 3(e), respectively.

\rev{\rev{This construction does not require a linear environment; any monotonic force--length relation $f_{\rm env}(l)$, such as the strain-stiffening response of fibrous collagen networks~\cite{Nam2021,Rabbi2026}, can replace Eq.~(\ref{eq:f-env}), with equilibrium again given by its intersection with $f_{\rm cell}(l)$.}}

\rev{In addition to the confinement-dependent force generated by cell deformation, active forces generated by the mitotic spindle can contribute to the axial mechanical balance~\cite{Nam2018,Gupta2021,Nam2021}. We denote by $F_\mathrm{spindle}$ the axial force generated by the spindle and take $F_\mathrm{spindle} > 0$ when it acts in the same direction as the compressive force exerted by the cell on the confining environment. The total force exerted by the cell on the environment is then
\begin{equation}
  f_\mathrm{env} = f_\mathrm{cell} + f_\mathrm{spindle},
  \label{eq:f-spindle}
\end{equation}
where $f_\mathrm{spindle} = F_\mathrm{spindle}/2\pi\gamma R_\mathrm{cell}$. A constant spindle force thus shifts the environmental force--length relation relative to the rigid-cell force--confinement relation, changing the operating point, as illustrated in Fig.~3(f). Equivalently, its effect can be incorporated into the spring description by redefining the reference length as $X'_\mathrm{ref} = X_\mathrm{ref} - F_\mathrm{spindle}/k$, or $\lambda' = \lambda + f_\mathrm{spindle}/\kappa$, so that equilibrium requires $f_\mathrm{cell}(l^*) = \kappa\,(l^* - \lambda')$. }

\rev{Increasing the spindle force therefore shifts the environmental force--length relation to larger $l$, and since $f_\mathrm{cell}(l)$ decreases with $l$, the operating point moves to a smaller $f^*$.
Here $f^*$ is the force exerted by the cell surface, which sets the mitotic pressure and the minimum furrow radius. The force on the environment, $f_{\rm env} = \kappa(l^* - \lambda)$, instead increases. Since $\rho_{\rm div}^{\min}$ decreases with $l$ (Fig.~2(f)), the spindle also allows further furrow ingression. Between rigid walls ($\kappa\to\infty$), $l^*=\lambda$ is fixed, so the spindle force adds to the force on the walls without changing the cell shape or the mitotic pressure. }

\section{Three-dimensional confinement}

\begin{figure}
    \centering
  \includegraphics[width=\linewidth]{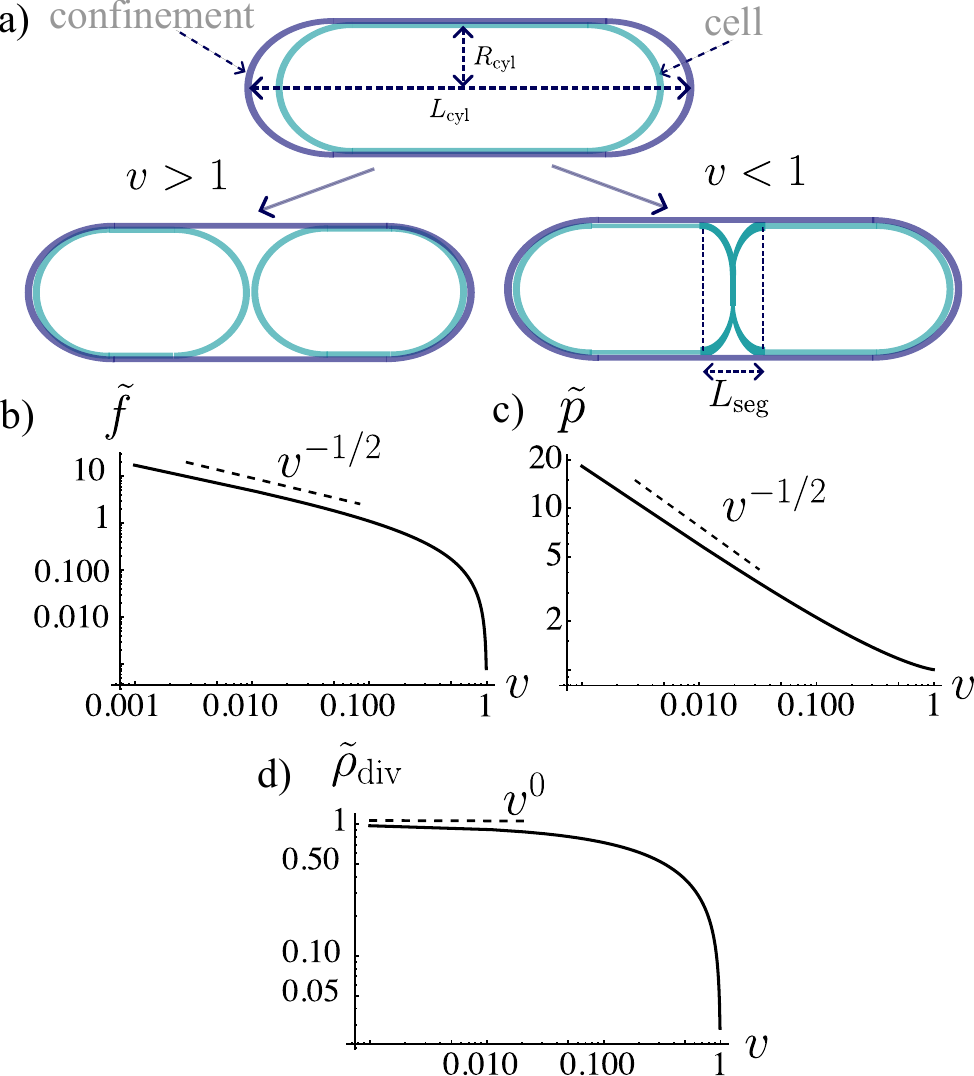}
   \caption{\textbf{Cell division under cylindrical confinement.}
(a) Geometry of a cell confined within a cylindrical chamber of radius $R_\mathrm{cyl}$ and total length $L_\mathrm{cyl}$, with hemispherical caps. Two regimes are shown: for $v\geq 1$, the cell accommodates division by axial elongation without transmitting axial force; for $v<1$, ingression requires pushing against the caps, leading to nonzero axial force. $L_\mathrm{seg}$ denotes the axial extent of the free segments around the division plane.
(b)--(d) Final state, in which the furrow has reached its minimum attainable radius, as a function of the dimensionless free volume $v=\Delta V/\Delta V_\mathrm{th}$, on logarithmic axes. Dashed lines indicate the strong-confinement behavior ($v\ll 1$).
(b) Dimensionless axial force $\tilde f$, with $\tilde f\sim v^{-1/2}$.
(c) Scaled pressure $\tilde p$, with $\tilde p\sim v^{-1/2}$. 
(d) Minimum furrow radius $\tilde\rho_{\rm div}^{\rm min}$, which approaches the chamber radius as $v \to 0$ (dashed line, $\tilde\rho = 1$). The approach follows $1 - \tilde\rho_{\rm div}^{\rm min} \sim v^{1/2}$. }
    \label{fig4}
\end{figure}

Thus far, confinement has been considered in the parallel-wall geometry, where the axial separation is prescribed while the radial extent of the cell remains unconstrained. We now extend this analysis to three dimensions, where both axial and radial deformations are restricted. 
\rev{We consider a cell dividing inside a rigid cylindrical chamber of radius $R_{\mathrm{cyl}}$ and total length $L_{\mathrm{cyl}}$, closed at both ends by hemispherical caps (Fig.~\ref{fig4}(a)). 
The total available volume inside the chamber is
\begin{equation}
\nonumber  V_\mathrm{cyl} = \pi R_\mathrm{cyl}^2 \left(L_\mathrm{cyl} - 2R_\mathrm{cyl}\right)
  + \frac{4}{3}\pi R_\mathrm{cyl}^3 .
\end{equation}
As in Sec.~II, the dividing cell has total volume $2V_\mathrm{cell} = 8\pi R_\mathrm{cell}^3/3$, where $R_\mathrm{cell}$ is the radius of a sphere with the volume of one daughter cell. 
Using this, we define the free volume as
\begin{equation}
  \Delta V = V_\mathrm{cyl} - 2V_\mathrm{cell},
\end{equation}
which quantifies the space available to accommodate shape changes during division.
We analyze the case when $R_\mathrm{cell} > R_\mathrm{cyl}$, so that each daughter cell, and hence the cell throughout division, is radially confined, and its maximum radius is fixed to $R_\mathrm{max} = R_\mathrm{cyl}$. For $R_\mathrm{cell} < R_\mathrm{cyl}$ the cells are not radially confined, and we do not consider this case here.}
\rev{The axial force transmitted along the division axis is controlled by this available free volume. As $\Delta V$ is reduced, axial elongation becomes progressively restricted, and once the cell surface comes into contact with the hemispherical caps, axial force transmission becomes unavoidable.
}

\rev{Cap contact occurs below a threshold free volume $\Delta V_{\mathrm{th}} = 2\pi R_\mathrm{cyl}^3/3$, the difference between the volume of a cylindrical section of length $2R_\mathrm{cyl}$ and that of two hemispheres of radius $R_\mathrm{cyl}$, which division must displace axially to form two cylindrical daughter cells with hemispherical ends.
We therefore define the dimensionless free volume
\begin{equation}
v\equiv \Delta V/\Delta V_{\mathrm{th}},
\end{equation}
so that $v=1$ marks the onset of cap contact and $v\ll1$ corresponds to strong three-dimensional confinement.}

\rev{Prior to contact with the hemispherical caps, the hydrostatic pressure is set by the free ends of the cell, whose radius of curvature is $R_\mathrm{cyl}$, so that by the Laplace law $\Delta P_\mathrm{cyl} = 2\gamma/R_\mathrm{cyl}$. 
If the cell contacts the cylinder over an axial length $L_\mathrm{contact}$, the normal stress on the wall is $\Delta P_\mathrm{cyl} - \gamma/R_\mathrm{cyl} = \gamma/R_\mathrm{cyl}$, since the contacting interface is curved only in the azimuthal direction. The integrated normal force on the cylindrical wall is therefore $F_\mathrm{rad}/F_\mathrm{cyl} = L_\mathrm{contact}/R_\mathrm{cyl}$, where $F_\mathrm{cyl} \equiv 2\pi\gamma R_\mathrm{cyl}$. In this regime, the axial compressive force on the hemispherical caps remains zero. }

For $v < 1$, the surface between the division plane and the cap contact point separates into a contact region, where the geometry is imposed by the confinement, and a free segment not in contact with the chamber. This free segment extends from the division plane ($r=R_\mathrm{div}$) to the detachment point from the cylinder and obeys the force-balance condition given by Eq.~\eqref{eq:forcebalance1}. 
\rev{Since the cell size enters only through the free volume, cells of different sizes in the same chamber follow the same force--free-volume relation, whereas different chamber radii give different relations. We therefore non-dimensionalize lengths by $R_{\rm cyl}$, forces by $F_{\rm cyl} = 2\pi\gamma R_{\rm cyl}$, and pressure by $\Delta P_{\rm cyl} = 2\gamma/R_{\rm cyl}$, and denote the dimensionless local radius, force, and pressure by $\tilde\rho = r/R_{\rm cyl}$, $\tilde f = F/F_{\rm cyl}$, and $\tilde p = \Delta P/\Delta P_{\rm cyl}$ (Table~\ref{tab:notation}). In terms  of these variables, Eq.~\eqref{eq:forcebalance1} becomes
\begin{equation}
\label{eq:forcebalance3}
\tilde\rho^2-\frac{\cos\theta}{\tilde p}\tilde\rho-\frac{\tilde f}{\tilde p}=0.
\end{equation}
}

\rev{To understand how the chamber geometry constrains division, we consider the final state of the dividing cell, in which the furrow has reached its minimum attainable radius $R_\mathrm{div}^\mathrm{min}$, so that further contraction of the cytokinetic ring no longer changes the cell shape. We ask what furrow radius, axial force, and mitotic pressure the chamber allows in this state.}

\rev{Evaluating the force balance at the detachment point by substituting
$\tilde \rho=1$ and $\theta=0$ in Eq.~\eqref{eq:forcebalance3}
gives
\begin{equation}
\tilde f=\tilde p-1.
\label{eq:fb-3d}
\end{equation}
The minimum furrow radius, obtained by substituting Eq.~\eqref{eq:fb-3d} in Eq.~\eqref{eq:min-rho}, reads
   \begin{equation}
     \tilde\rho^{\rm min}_{\rm div} = \sqrt{(\tilde p - 1)/\tilde p}.
     \label{eq:rhomin-3d}
   \end{equation}
The pressure $\tilde p$ is determined by enforcing volume conservation for the free segment.    For a given $\tilde p$, Eqs.~\eqref{eq:forcebalance3}-\eqref{eq:rhomin-3d} determine the free segments between $\tilde\rho = 1$ and $\tilde\rho = \tilde\rho_{\rm div}^{\rm min}$, and hence their total axial  extent $\tilde L_\mathrm{seg}$ and enclosed volume $\tilde V_\mathrm{seg}$, scaled by $R_\mathrm{cyl}$ and $R_\mathrm{cyl}^3$ and counting both sides of the division plane. The pressure then follows from
\begin{equation}
    \pi \tilde L_{\mathrm{seg}} - \tilde V_{\mathrm{seg}} = \frac{2\pi}{3}\,v.
    \label{eq:excess-vol}
\end{equation}
}
\rev{The calculation thus parallels that of Sec.~\ref{sec:rigid}, with $R_\mathrm{cyl}$ in place of $R_\mathrm{cell}$. The only difference lies in the constraint that closes the problem: under parallel-wall confinement, $f$ and $p$ are fixed by the wall separation and the cell volume, whereas here tangential detachment from the cylinder wall relates them through Eq.~\eqref{eq:fb-3d}, leaving $\tilde p$ to be fixed by the free volume alone. The free volume $v$ therefore plays the role of the wall separation $l$, and the dimensionless final state depends only on $v$, independent of the absolute cell size and surface tension, which enter only through $v$ and the scales $F_\mathrm{cyl}$ and $\Delta P_\mathrm{cyl}$.}

\rev{The force, pressure, and minimum furrow radius in the final state are shown as functions of $v$ in Figs.~\ref{fig4}(b), \ref{fig4}(c), and \ref{fig4}(d), respectively. At $v = 1$, $\tilde p = 1$, so that $\tilde f = 0$ and $\tilde\rho_\mathrm{div}^\mathrm{min} = 0$; as $v$ decreases, the force and pressure increase, and the minimum furrow radius approaches the confinement radius.}

\rev{In the strong-confinement regime, $v \ll 1$, the dependence on free volume follows a simple geometric scaling. The small free volume forms a region whose radial and axial extent is set by the curvature radius of the free segment, $R_\mathrm{cyl}/\tilde p$, giving $\Delta V \sim R_\mathrm{cyl}^3/\tilde p^{\,2}$, and hence $\tilde p \sim v^{-1/2}$. Eq.~\eqref{eq:fb-3d} then gives $\tilde f \sim v^{-1/2}$, and Eq.~\eqref{eq:rhomin-3d} gives $1 - \tilde\rho_\mathrm{div}^\mathrm{min} \sim v^{1/2}$. These power laws are evident in the logarithmic plots in Fig.~\ref{fig4}. }

\rev{The radial force on the cylindrical wall is $F_{\rm rad}/F_{\rm cyl}=(2\tilde p-1)\tilde L_{\rm contact}$. For $v\ge 1$, $\tilde p=1$, so the force per unit contact length is fixed and $F_{\rm rad}$ changes only through $L_{\rm contact}$. For $v<1$, the pressure also rises, and $\tilde L_{\rm contact}=\tilde L_{\rm cyl}-2-\tilde L_{\rm seg}$. Since $\tilde L_{\rm seg}\sim v^{1/2}$, the contact length approaches a constant at strong confinement, and the radial force grows as $v^{-1/2}$, like the axial force.}

\rev{The construction used here is not specific to the cylinder and applies to any axisymmetric enclosure. Let the enclosure be described by its profile $\rho_e(z)$. In the confinement-limited state, the cell leaves the enclosure at some position $z_w$, where its radius is $\rho_w = \rho_e(z_w)$ and its tangent angle is $\theta_w$. For a non-adhesive enclosure, the cell leaves tangentially, so $\theta_w$ equals the angle of the wall, $\tan\theta_w = \rho_e'(z_w)$. With adhesion, $\theta_w$ also includes the contact angle set by the adhesion energy, as for the septum in Sec.~II. Evaluating Eq.~(2) at this point gives the force,
\begin{equation}
f = p\,\rho_w^2 - \rho_w\cos\theta_w,
\label{eq:f_general}
\end{equation}
which reduces to Eq.~(12) for a non-adhesive cylinder, where $\theta_w = 0$ and $\rho_w = 1$ in units of $R_{\rm cyl}$. The free segment, which starts at the furrow with $\theta = \pi/2$ and radius $\sqrt{f/p}$, must reach the enclosure at $z_w$. Together with the free volume, this fixes $p$ and $z_w$. Each enclosure geometry therefore yields its own constitutive relations, $f(v)$, $p(v)$ and $\rho_{\rm div}^{\min}(v)$, in terms of the free volume.}

\section{Inferring division mechanics from cell shape}
\label{sec:readout}

\rev{The parallel-wall geometry of Sec.~II is realized mainly in vitro, between plates or under a cantilever. In tissues and gels, by contrast, dividing cells are enclosed on all sides by a deformable environment. Here we ask what can be inferred about the mechanics of division from the geometry of the cell alone, without knowing the material properties of the enclosure.}

\rev{In the confinement-limited state, the part of the cell not in contact with the enclosure obeys the force balance of Eq.~(1), and its tangent angle is known at both ends. At the stalled furrow, of radius $R_{\rm div}^{\min}$, the tangent is perpendicular to the division axis. At the point where the cell leaves a non-adhesive enclosure, of radius $R_w$, the cell is tangent to the wall, so its angle equals that of the wall, $\theta_w$. Evaluating Eq.~(1) at the two ends and solving for pressure and force gives
\begin{equation}
\frac{\Delta P}{\gamma} = \frac{2R_w\cos\theta_w}{R_w^2 - \left(R_{\rm div}^{\min}\right)^2}, \qquad \frac{F_{\rm cell}}{\gamma} = \frac{2\pi R_w\left(R_{\rm div}^{\min}\right)^2\cos\theta_w}{R_w^2 - \left(R_{\rm div}^{\min}\right)^2}.
\label{eq:F_readout}
\end{equation}
Where the wall is parallel to the division axis at detachment, as in a cylinder, $\theta_w = 0$, and two radii suffice; Eqs.~(12) and (13) are then recovered for $R_w = R_{\rm cyl}$. For an axisymmetric cell, these quantities can be measured on a single section through the division axis, and the geometry thus determines pressure and force in units of the tension. The force exerted on the environment is $F_{\rm env} = F_{\rm cell} + F_{\rm spindle}$ (Sec.~III). If $F_{\rm env}$ is measured independently from the mechanics of the surroundings, Eq.~(\ref{eq:F_readout}) gives $F_{\rm spindle} = F_{\rm env} - F_{\rm cell}$, provided $\gamma$ is known.}

\rev{Equation~(\ref{eq:F_readout}) relies only on axial force balance at the two ends of the free segment, and therefore extends directly to more general conditions. With adhesion to the enclosure, $\theta_w$ is replaced by the angle of the free membrane, and  with an adhesive septum, the furrow angle is replaced by contact angle. For spatially varying, anisotropic and time-dependent surface stresses, Eq.~(1) holds at each instant with $\gamma$ replaced by the local meridional stress $T_s$ (Appendix~\ref{appendix_general}). The geometry then gives the instantaneous pressure and force, provided the furrow is at its instantaneous minimum radius and pressure gradients within the cytoplasm are negligible.}

\rev{A quantitative test of these relations requires measurements that existing experiments do not report. Studies of division in gels and monolayers measure the elongation of the cell and the deformation of its surroundings~\cite{Nam2018,Gupta2021,Nam2021}, and infer forces through a constitutive model of the matrix. The theory instead requires the radius and tangent angle of the cell at the furrow and where it leaves its surroundings, and imaging of the ring separately from the membrane to identify when the furrow stalls. We therefore restrict the comparison to orders of magnitude. For $\gamma \approx 1$--$2\,\mathrm{mN/m}$~\cite{Fischer2014} and $R_{\rm cyl} \approx 8$--$10\,\mu\mathrm{m}$, moderate confinement, $v \approx 0.1$ or $\tilde p \approx 2$--$3$, gives pressures of $0.4$--$1.5\,\mathrm{kPa}$ and, from Eq.~(\ref{eq:fb-3d}), forces of $50$--$250\,\mathrm{nN}$. These values are comparable to the protrusive stresses of $1$--$2$\,kPa measured during late division in alginate gels~\cite{Nam2018}. The forces of $300$--$500$\,nN measured in epithelial monolayers~\cite{Gupta2021} exceed this range and point to a spindle contribution or stronger confinement. A quantitative comparison requires modeling viscoelastic and plastic response of the confining matrices, along with imaging of the cell shape, which is left for future work.}

\section{Discussion}

We have developed a minimal mechanical theory of how confinement limits the shapes a dividing cell can adopt, and hence the pressure and force it can transmit. Confinement imposes a minimum attainable furrow radius. Once the cytokinetic ring reaches it, further contraction no longer changes the interface, so mitotic pressure and transmitted force saturate while closure continues. This separates the ring radius, an active input, from the furrow radius, a geometric outcome. The three geometries share this mechanism and differ in what sets the bound: the wall separation in rigid confinement, the operating point $(l^*, f^*)$ selected by environmental stiffness in soft confinement, and the free volume $v$ in cylindrical confinement.

The confinement-limited state offers an alternative route to septum formation. In Ref.~\cite{Sain2015}, a septum forms when its tension is lower than that of the outer surface, for example through adhesion. Under confinement, a septum of outer radius $R_{\rm div}^{\min}$ can form even when the two tensions are equal. In the \textit{C.~elegans} embryo, which divides inside a rigid eggshell, the ring constricts while separating from the furrow, leaving a septum of fixed outer radius~\cite{Carvalho2009,Zumdieck2007,Sain2015}. In cells isolated from the eggshell, the membrane furrow instead closes~\cite{Koyama2012}, and no septum forms~\cite{Sain2015}. 

The mechanical output of a division depends on the geometry, mechanics, and adhesiveness of the environment within which the cell deforms. This is relevant to continuum descriptions of tissues, where divisions are often represented as force dipoles~\cite{Ranft2010,Tahaei2025,Godard2019}. A cell pushing against a deformable environment transmits a force that depends on the stiffness of the environment and on adhesion to it, so the effective dipole strength of a division need not be a fixed property of the cell. In three-dimensional tissues, the free volume is set by the deformation of the surrounding tissue. The constitutive relation of the corresponding enclosure, combined with the response of the tissue, then determines the force and the free volume, as for soft axial confinement.

\rev{Simulations accompanying recent force measurements~\cite{Gupta2021,Nam2021,Rabbi2026} rest on the same volume-conservation mechanism. They represent the cell surface as an elastic membrane and impose ring constriction on equatorial membrane nodes, so that the transmitted force depends on membrane elastic parameters and the ring and furrow cannot separate. Treating the mitotic cortex instead as a constant-tension interface, as supported by measurements on mitotic cells~\cite{Fischer2014}, yields the pressure, force bounds and minimum furrow radius analytically. Models of the active cortex address how active stresses and flows produce the furrow without confinement~\cite{Turlier2014,Rajput2025,Mukherjee2026}, while work that includes the environment has focused on division dynamics and stability~\cite{Plan2020,Cuvelier2023,Gao2026}. Our approach complements them by isolating the geometric limit that confinement places on furrow ingression, force and pressure.}

\rev{The theory rests on several simplifications. We treat the cortex as an interface of constant tension, but the minimum furrow radius does not depend on this assumption. It persists for spatially varying, anisotropic and time-dependent surface stresses, and the pressure and force inferred from cell shape are then instantaneous values (Appendix~B). Adhesion enters only through contact angles at the septum and at the enclosure. We neglect re-adhesion of the daughters to the substrate~\cite{Stewart2011} and the nonlinear, viscoelastic response of fibrous matrices.}

\rev{We also restrict to shapes that are axisymmetric and symmetric about the division plane. Ingression can become asymmetric at high contractility~\cite{Rajput2025}, and constant-tension interfaces are prone to bubble-type instabilities~\cite{Cuvelier2023}. The minimum furrow radius therefore applies only if the symmetric shape is attained, which active positioning of the ring may help ensure~\cite{Green2012}. We assume quasistatic equilibrium and constant volume. Slow volume changes~\cite{Gupta2021,Cadart2014,Zlotek-zlotkiewicz2015} would shift the force--confinement relation but would not remove the minimum furrow radius. Finally, apart from the example of septum closure in Appendix~C, the theory gives the mechanical state for a prescribed ring radius rather than the time course of cytokinesis.}

\rev{The clearest experimental test of the theory is the separation of the cytokinetic ring from the outer furrow, which can be identified by imaging the ring and the membrane separately. Once the furrow has stalled, its shape gives the mitotic pressure and transmitted force without a model of the surroundings. Natural extensions of this work are division in deformable three-dimensional environments, where the constitutive relation of the enclosure must be combined with the viscoelastic and plastic response of the matrix, and the coupling of the geometric limit to active descriptions of the cortex and the ring.}

\section{Acknowledgments}
 
ASV acknowledges funding from Leverhulme Trust (LIP-2021-017).

\appendix

\section{Shape under confinement}
\label{appendix_shape}

We derive the shape of a confined dividing cell assuming axisymmetry and constant surface tension. The cell surface is modeled as a constant--mean--curvature (CMC) surface enclosing an incompressible fluid \cite{Brochard2003,Fischer2014,Yang2017,Vishen2024}.

The surface is parametrized by arc length $s$ and local scaled radius $\rho(s)$, with $\theta(s)$ the angle between the surface tangent and the axis of symmetry. We choose $s=0$ at the point of maximum radius $\rho=\rho_{\max}$, where $\theta=0$. 

The division plane (furrow) and the confining wall correspond to arc-length positions $s = s_d$ and $s = s_w$, respectively, on opposite sides of $s = 0$. At the division plane, $\rho(s_d) = \rho_\mathrm{div}$. We take the wall to be non-adhesive, so that the interface meets it tangentially, $\theta(s_w) = \pi/2$; from Eq.~(2), the radius of the wall contact region is then $\rho_w = \sqrt{f/p}$. Since the same condition $\theta = \pi/2$ holds at the division plane in the confinement-limited state, the minimum furrow radius equals the radius of the wall contact region.

From Eq.~\eqref{eq:forcebalance2},
\begin{equation}
\cos\theta = \rho ~p - \frac{f}{\rho},
\label{eq:cos_theta}
\end{equation}
and using $d\rho/ds=\sin\theta$, we obtain
\begin{equation}
\left(\frac{d\rho}{ds}\right)^2
= 1 - \left(\rho~p - \frac{f}{\rho}\right)^2
= \frac{p^2(\rho_{\max}^2-\rho^2)(\rho^2-\rho_{\min}^2)}{\rho^2},
\label{eq:shape_ode}
\end{equation}
with turning points
\begin{equation}
\rho_{\max,\min}
= \frac{1}{2 p}
\left(1 \pm \sqrt{1+4 f~p}\right).
\end{equation}
For compressive confinement ($ f>0$), $\rho_{\min}<0$ and does not represent a physical radius. It is retained as a formal parameter that simplifies the integration.

Introducing the dimensionless variable $u=s ~p$ and the parameter
\begin{equation}
\alpha = 1-\left(\frac{\rho_{\min}}{\rho_{\max}}\right)^2,
\end{equation}
Eq.~\eqref{eq:shape_ode} integrates to the standard CMC profile
\begin{equation}
\rho(s)=\rho_{\max}\sqrt{1-\alpha\sin^2 u}.
\label{eq:r_of_s}
\end{equation}
The axial coordinate follows from $dz/ds=\cos\theta$. Substituting
Eq.~\eqref{eq:cos_theta} and integrating yields
\begin{equation}
z(s)
= \frac{1}{p}\left[
\rho_{\max}~p \,\mathbb{E}(u,\alpha)
-\frac{f}{\rho_{\max}}\,\mathbb{F}(u,\alpha)
\right],
\label{eq:z_of_s}
\end{equation}
where $\mathbb{F}$ and $\mathbb{E}$ are the incomplete elliptic integrals of the first and second kind, respectively, and $z(0)=0$ at $\rho=\rho_{\max}$.

The arc-length positions $s_d$ and $s_w$ are obtained from Eq.~(A5) by setting $\rho = \rho_\mathrm{div}$ and $\rho = \rho_w$, respectively. The axial extent of one daughter cell, from the division plane to the wall, is
\begin{equation}
  2L = |z(s_d)| + |z(s_w)|,
\end{equation}
so that the total axial extent is $4L$, consistent with Fig.~1. The volume of one daughter cell, enclosed between the division plane and the wall, is $V_\mathrm{cell} = |V(s_d)| + |V(s_w)|$, where
\begin{equation}
  \frac{V(s)}{R_\mathrm{cell}^3} = \pi \int_0^s \rho^2(s')
  \frac{dz}{ds'}\, ds'.
\end{equation}
This integral can be expressed in closed form using elliptic integrals.
For a prescribed furrow radius $\rho_{\rm div}$, the force $f$ and pressure $p$ are determined self-consistently by the two constraints, the wall separation, Eq.~(A7), and the cell volume, Eq.~(A8).

\rev{\section{Axial force balance for a general axisymmetric surface}
\label{appendix_general}
}
\rev{We consider a general axisymmetric interface supporting
an anisotropic and spatially varying surface stress. Following
the surface-traction formulation of membrane mechanics
\cite{Salbreux2017}, the force transmitted
across a curve $\partial S$ on the surface is obtained by
integrating the surface traction along the curve,
\begin{equation}
\mathbf F_{\mathrm{surf}}
=
\oint_{\partial S}
\boldsymbol{\sigma}_s\cdot\boldsymbol{\nu}\,dl,
\end{equation}
where $\boldsymbol{\sigma}_s$ is the surface stress tensor
and $\boldsymbol{\nu}$ is the unit co-normal to the curve,
lying in the tangent plane of the surface.}

\rev{For an axisymmetric surface, we introduce the meridional and
azimuthal directions, $\mathbf e_s$ and $\mathbf e_\phi$,
respectively. In the absence of shear stresses, the surface
stress tensor can be written as
\begin{equation}
\boldsymbol{\sigma}_s
=
T_s(s,t)\,\mathbf e_s\mathbf e_s
+
T_\phi(s,t)\,\mathbf e_\phi\mathbf e_\phi,
\end{equation}
where $T_s$ and $T_\phi$ denote the meridional and azimuthal
stresses. Consider a circular contour at fixed meridional
coordinate $s$. Its co-normal is $\boldsymbol{\nu}=\mathbf e_s$,
and therefore
\begin{equation}
\boldsymbol{\sigma}_s\cdot\boldsymbol{\nu}
=
T_s(s,t)\mathbf e_s.
\end{equation}
Integrating around the circumference of radius $r(s)$ and projecting onto the symmetry axis $\mathbf e_z$ gives
\begin{equation}
F_z
=
2\pi r(s)T_s(s,t)\cos\theta,
\label{eq:general_axial_force}
\end{equation}
where $\mathbf e_s\cdot\mathbf e_z=\cos\theta$, with $\theta$ defined as the angle between the meridional tangent and the axis of symmetry.}

\rev{Thus, the axial force transmitted across a circular contour depends on the local meridional stress, whereas the azimuthal stress does not contribute directly to this force. The latter can nevertheless affect the shape and stress distribution through the local force balances.
The axial force balance, including pressure and other applied forces, consequently provides a direct relation between the axisymmetric shape and the meridional surface stress. It can be used either to determine the shape for a prescribed meridional stress distribution or, conversely, to constrain the meridional stress for a prescribed shape.}

\rev{Using the same dimensionless force and pressure scales as in the main text, defining the dimensionless meridional stress as $\tilde T_s = T_s/\gamma$, and making the same approximations as in Sec.~II (uniform cytoplasmic pressure and quasistatic force balance), the force balance becomes
\begin{equation}
\rho_{\mathrm{div}}^2
-
\frac{\tilde T_s(s_{\mathrm{div}},t)\cos\theta}{p(t)}
\rho_{\mathrm{div}}
-
\frac{f(t)}{p(t)}
=0.
\label{eq:general_shape}
\end{equation}
The minimum radius is attained at the division plane for $\theta=\pi/2$, that reads 
\begin{equation}
\rho_{\mathrm{div}}^{\min}(t)
=
\sqrt{\frac{f(t)}{p(t)}}.
\label{eq:general_minimum}
\end{equation}
The generalized axial force balance therefore predicts a finite minimum furrow radius even when the applied force and pressure are time-dependent. Unlike the constant-tension case, the limiting radius need not remain fixed, but is determined instantaneously by the ratio of the applied force to the pressure difference. These quantities are constrained by volume conservation and the imposed confinement geometry, and may continue to evolve as the mechanical state changes. Consequently, the existence of a minimum furrow radius does not require saturation of the force or pressure. Rather, confinement imposes a geometric lower bound on furrow ingression, whose value can change as the force and pressure evolve. In the absence of compressive loading (\(f=0\)), this confinement-induced lower bound vanishes.}

\rev{\section{Ring closure in the confinement-limited state}
\label{appendix_ring}
}
\rev{The analysis in the main text determines the quasistatic outer shape, force and pressure for a prescribed ring radius, and does not specify how the ring contracts. Once the furrow reaches $R_{\rm div}^{\min}$, the outer shape is fixed and further ring contraction proceeds within a septum whose outer radius is $R_{\rm div}^{\min}$. Any ring dynamics that treats the septum as a domain of fixed outer radius, with vanishing velocity of the septum material at its outer edge, is therefore compatible with the present framework, provided dynamic contributions to the septum stress at its outer edge remain small compared with the static septum tension, so that the furrow angle remains at $\theta=\pi/2$. The septum-closure model of Ref.~\cite{Sain2015}, in which ring constriction is balanced by septum tension, septum viscosity and ring friction under exactly this boundary condition, applies with outer septum radius $R_{\rm div}^{\min}$ and septum tension $2\gamma$, or $2\gamma_s$ for a septum of reduced tension (Sec.~II). Active-gel descriptions in which the ring forms within an actomyosin layer on a septum of fixed outer radius~\cite{Chatterjee2022} can be treated in the same way. In this regime, ring contraction is decoupled from the outer cell shape, and the force and pressure transmitted to the confinement remain constant while the ring closes.}

\rev{The framework does not apply when cortical flows persist on the outer cell surface, outside the septum, as in descriptions of furrowing driven by gradients of cortical contractility~\cite{Turlier2014}. Such flows render the surface stress non-uniform and anisotropic, so the outer shape is no longer a constant-tension equilibrium and continues to evolve.  In that case the minimum furrow radius is still bounded by the instantaneous force and pressure (Appendix~\ref{appendix_general}), but it need not remain fixed.}

\rev{As an example, consider the symmetric septum closure of Ref.~\cite{Sain2015}, in the limit where friction within the ring dominates over septum viscosity. The ring, of radius $R_{\rm cyto}$, line tension $\Sigma_0$ and internal friction $\zeta_L$~\cite{Sain2015}, is pulled outward by the septum tension $2\gamma$. Radial force balance at the inner septum edge gives
\begin{equation}
\frac{\zeta_L}{R_{\rm cyto}}\frac{dR_{\rm cyto}}{dt} = 2\gamma R_{\rm cyto} - \Sigma_0 .
\end{equation}
With $x = R_{\rm cyto}/R_{\rm div}^{\min}$, $\tau_L = \zeta_L/\Sigma_0$, and $\tilde \sigma = 2\gamma R_{\rm div}^{\min}/\Sigma_0$, and starting from $x = 1$ at the time $t^*$ when the minimum furrow radius is reached, the solution is
\begin{equation}
x(t) = \frac{1}{(1-\tilde \sigma)\, e^{(t-t^*)/\tau_L} + \tilde\sigma} .
\end{equation}
In this example, closure therefore requires $\tilde\sigma < 1$, i.e.\ $\Sigma_0 > 2\gamma R_{\rm div}^{\min}$. Stronger confinement, which increases $R_{\rm div}^{\min}$, raises the ring tension needed to complete cytokinesis. At late times $x \propto e^{-(t-t^*)/\tau_L}$, set by ring friction alone. Throughout this stage the outer cell shape, and hence the force and pressure transmitted to the confinement, remain constant.}

	\bibliographystyle{apsrev4-2}
	\bibliography{CellDivision}{}
	
\end{document}